\documentclass[a4paper,11pt]{llncs}
\usepackage{pure-report}
\usepackage{amsmath}
\usepackage{fancyvrb}
\usepackage{a4wide}

\usepackage{mymacros}

\DeclareRobustCommand\dcolon{\mathbin{::}}

\def\titulo{Deriving Sorting Algorithms}
\title{\titulo}
\author{Jos\'e Bacelar Almeida and Jorge Sousa Pinto\\
Departamento de Inform\'atica\\ 
Universidade do Minho\\
4710-057 Braga, Portugal\\
\email{\{jba,jsp\}@di.uminho.pt}
}
\def\autor{Jos\'e Bacelar Almeida and Jorge Sousa Pinto}
\def\data{2006, April}
\def\numero{DI-PURe-06.04.01}
\def\resumo{This paper shows how 3 well-known sorting algorithms can
  be derived by similar sequences of transformation steps from a
  common specification. Each derivation uses an auxiliary algorithm
  based on insertion into an intermediate structure. The proofs given
  involve both inductive and coinductive reasoning, which are here
  expressed in the same program calculation framework, based on
  unicity properties.}

\begin{document}
\purecover

\bibliographystyle{plain} 

\title{\titulo}

\author{Jos\'e Bacelar Almeida and Jorge Sousa Pinto\\
\email{\email{\{jba,jsp\}@di.uminho.pt}}}
\institute{Departamento de Inform\'atica\\ 
Universidade do Minho\\
4710-057 Braga, Portugal
}
\date{\data}

\maketitle 

\begin{abstract}
This paper proposes new derivations of three well-known sorting
algorithms, in their functional formulation. The approach we use is
based on three main ingredients: first, the algorithms are derived
from a simpler algorithm, i.e.  the specification is already a
solution to the problem (in this sense our derivations are program
transformations). Secondly, a mixture of inductive and coinductive
arguments are used in a uniform, algebraic style in our
reasoning. Finally, the approach uses structural invariants so as to
strengthen the equational reasoning with logical arguments that cannot
be captured in the algebraic framework.
\end{abstract}


\section{Introduction}

This paper presents new derivations of three well-known sorting
algorithms, in the functional setting. 
Our approach can be summarized as follows:

\begin{enumerate}

\item It is based on \emph{program transformation} in the sense that
  we depart from a specification that is already a (not very
  efficient) algorithm for solving the problem. Traditional
  derivations of sorting algorithms (building on the work of Burstall
  and Darlington) formalize the ``is sorted'' property on
  lists. Instead, we take the insertion sort algorithm to be a
  specification of sorting, and derive, by sequences of correct steps,
  more efficient algorithms from it.

\item The algorithms that we derive follow the \emph{derive and
    conquer} strategy and as such are not structurally recursive on
  their arguments. For this reason a combination of inductive and
  coinductive reasoning must be used. We adhere here to the equational
  style of reasoning usually known to functional programmers as
  \emph{program calculation}, which relies on uniqueness properties of
  certain recursion patterns. Although the proofs are independent of
  this choice, we find that this allows for greater uniformity between
  the inductive and coinductive arguments.

\item In two of our three derivations, the equational reasoning must
  be strengthened by using invariants on certain intermediate
  data-structures, since some of the equalities one needs to prove are
  not universal for a given data-type. For instance, it is not true
  that the in-order traversal of any binary tree produces a sorted
  list. This is however true for trees produced in a certain way. As
  far as we know there is very little work on program calculation
  strengthened with invariants.

\item The algorithms are derived as \emph{hylomorphisms}, i.e. as
  explicit compositions of a recursive function with a co-recursive
  one, with an intermediate data-structure of a tree type, which can
  be \emph{deforested} to produce the standard formulation of the
  algorithms. Sorting algorithms have been defined as hylomorphims
  elsewhere~\cite{AugusteijnL:sorm}.

\end{enumerate}



The paper is structured as follows: Section~\ref{sec:sorting} reviews
standard material on sorting in the functional setting, including the
algorithms that will be considered in the main sections of the
paper. Section~\ref{sec:bacsah} contains background material on
program calculation, based on unicity (or universal) properties of
recursion pattern operators. Section~\ref{sec:sori} then introduces
two generic algorithms for sorting, based on insertion into an
intermediate structure of a container type (in a leftwards and
rightwards fashion
respectively). Sections~\ref{sec:mergesort},~\ref{sec:deriv-heap-sort},
and~\ref{sec:deriv-quick-sort} present the derivations of merge sort,
quicksort, and heapsort, which are based on instantiations of the
generic algorithms. Finally we conclude the paper in
Section~\ref{sec:con}.

\section{Sorting Homomorphisms and Divide-and-conquer Algorithms }        

\label{sec:sorting}

Consider a very simple algorithm for sorting a list, usually known
under the name of \emph{insertion sort}. We give it here written in
Haskell.
\begin{code}
isort []     = []
isort (x:xs) = insert x (isort xs) 
\end{code}
where \verb+insert+ inserts an element in a sorted list. This is
certainly a natural way of sorting a list in a traditional functional
language: since the list structurally consists of a head element
\texttt{x} and a tail sublist \texttt{xs}, it is natural to
recursively sort \texttt{xs} and then combine this sorted list with
\texttt{x}. This pattern of recursion can be captured by the
\tf{foldr} operator, resulting in the following definition where
explicit recursion has been removed.
\begin{code}
isort = foldr insert []
\end{code}
Actually, any sorting function is a \emph{list
  homomorphism}~\cite{BirdR:inttl,GibbonsJ:thiht}, which means that if
the initial unsorted list is split at any point and the two resulting
sublists are recursively sorted, there exists a binary operator $\odot$
that can combine the two results to give the final sorted list.
$$
\f{isort}\ (l_1 \app l_2) = (\f{isort}\ l_1) \odot (\f{isort}\ l_2)
$$
This $\odot$ operator is of course the (linear time) function of type
$[a] \to [a] \to [a]$ that merges two sorted lists:

\begin{code}
merge [] l = l
merge l [] = l
merge (h1:t1) (h2:t2) 
      | (h1<=h2)  = h1:(merge t1 (h2:t2))
      | otherwise = h2:(merge (h1:t1) t2)
\end{code}
%
%
%
The operator $\odot$ is associative with the empty list as unit,
forming a monoid over lists.  It is also commutative. \tf{insert} can
be defined in terms of $\odot$ as follows
\begin{align}
\label{eq:inswrap1}  \f{insert}\ x\ l & = [x] \odot l
\end{align}

Insertion sort runs in quadratic time.  Most well-know efficient
sorting algorithms perform recursion \emph{twice}, on subsequences
obtained from the input sequence, and then combine the results (for
this reason they are called \emph{divide-and-conquer} algorithms). As
such, they do not fit the simple iteration pattern captured by
\tf{foldr}.
In the following we describe three different divide-and-conquer
algorithms.  

\paragraph{Heapsort.} The principle behind heapsort is to traverse the
list to obtain, in linear time, its minimum element $y$ and a pair of
lists of approximately equal size, containing the remaining elements
(function \tf{haux}). The lists are then recursively sorted and merged
together, and $y$ pasted at the head of the resulting list.
\begin{code}
haux x [] = (x,[],[])
haux x (y:ys) = let (z,l,r) = haux y ys
                in if x<z then (x,z:r,l) else (z,x:r,l)

hsort []     = []
hsort (s:xs) = let (y,l,r) = haux x xs
               in y:(merge (hsort l) (hsort r))
\end{code}

\paragraph{Quicksort.} The criterion for obtaining the two sublists is
here to use the head of the list (the only element accessible in
constant time) as a pivot used to separate the remaining elements. The
two sorted results need only be concatenated (with the pivot in the
middle) to give the final result. 

\begin{code}
qaux _ []    = ([],[])
qaux x (h:t) = let (l,r) = qaux x t
               in if h<=x then (h:l,r) else (l,h:r)

qsort []     = []
qsort (x:xs) = let (l,r) = qaux x xs 
               in (qsort l) ++ x:(qsort r)
\end{code}

\paragraph{Merge Sort.} This is similar to heapsort except that the
minimum element is not extracted when the list is traversed. For this
reason an extra base case is used.
\begin{code}
maux []     = ([],[])
maux (x:xs) = (x:b,a)  where (a,b) = maux xs

msort []  = []
msort [x] = [x]
msort xs  = let (l,r) = maux xs
            in merge (msort l) (msort r)
\end{code}

These functional versions of the algorithms may be difficult to
recognize for a reader used to the imperative formulations, where the
sorting is usually done in place, on indexed arrays. All three are
however widely known in the functional programming community
formulated as above.

\section{Recursion Patterns, Unicity, and Hylomorphisms}
\label{sec:bacsah}

We direct the reader to~\cite{GibbonsJ:calfp} for an extensive
introduction to the field of program calculation, and include here
only the basic notions needed for expressing the proofs included in
the paper.

The fold recursion pattern can be generalized for any regular type; in
the context of the algebraic theory of data-types folds are
\emph{datatype-generic} (in the sense that they are parameterized by
the base functor of the type), and usually called
\emph{catamorphisms}.  The result of a fold on a node of some tree
data-type is a combination of the results of recursively processing
each subtree (and the contents of the node, if not empty).

The dual notion is the \emph{unfold} (also called \emph{anamorphism}):
a function that constructs (possibly infinite) trees in the most
natural way, in the sense that the subtrees of a node are recursively
constructed by unfolding.

In the present paper we will need to work with two flavours of binary
trees: leaf-labelled (for merge sort) and node-labelled trees (for the
remaining algorithms). These types, and the corresponding recursion
patterns, are defined in Table~\ref{tab:trees}.

\begin{table}[tb]
\hrule
\begin{code}
data BTree a = Empty | Node a (BTree a) (BTree a)
type Heap    = BTree
data LTree a = Leaf (Maybe a) | Branch (LTree a) (LTree a)


unfoldBTree :: (b -> (Either (a,b,b) ())) -> b -> BTree a
unfoldBTree g x = case (g x) of 
            Right ()     -> Empty
            Left (y,l,r) -> Node y (unfoldBTree g l) (unfoldBTree g r)

foldBTree :: (a -> b -> b -> b) -> b -> BTree a -> b
foldBTree f e Empty        = e
foldBTree f e (Node x l r) = f x (foldBTree f e l) (foldBTree f e r)


unfoldLTree :: (b -> (Either (b,b) (Maybe a))) -> b -> LTree a
unfoldLTree g x = case (g x) of
            Right y    -> Leaf y
            Left (l,r) -> Branch (unfoldLTree g l) (unfoldLTree g r)

foldLTree :: (b->b->b) -> ((Maybe a)->b) -> LTree a -> b
foldLTree f e (Leaf x)     = e x
foldLTree f e (Branch l r) = f (foldLTree f e l) (foldLTree f e r)
\end{code}
\hrule
\caption{Types and recursion patterns for binary trees}


\label{tab:trees}
\end{table}

In principle, a fold is a recursive function whose domain is a type
defined as a least fixpoint (an initial algebra), and an unfold is a
recursive function whose codomain is defined as a greatest fixpoint (a
final coalgebra).  However, in lazy languages such as Haskell, least
and greatest fixpoints coincide, and are simply called recursive
types.

At an abstract level, folds (as well as other structured forms of
recursion, such as primitive recursion) enjoy an initiality property
among the algebras of the base functor of the domain type. In concrete
terms, this makes possible the use of \emph{induction} as a proof
technique. Dually, unfolds are final coalgebras; techniques for
reasoning about unfolds include \emph{fixpoint induction} and
\emph{coinduction}~\cite{GibbonsJ:promscp}.

\paragraph{Unicity.}

The \emph{program calculation} approach is based on the use of
initiality and finality directly as an equational proof
principle. Both properties can be formulated in the same framework, as
\emph{universal} or \emph{unicity} properties. In this paper we
generally adhere to the equational style for proofs, but often resort
to induction for the sake of simplicity (in particular when none of
the sides of the equality one wants to prove is directly expressed
using a recursion pattern, applying a unicity property may require
substantial manipulation of the expressions). See~\cite{PintoJS:poipt}
for a study of program calculation carried out purely by using fusion,
including an adequate treatment of strictness conditions.

We give below the unicity properties that we shall require in the rest
of the paper, for the \tf{foldr}, \tf{unfoldLTree}, and
\tf{unfoldBTree} operators.  A weaker \emph{fusion} law for \tf{foldr}
is also shown, which is easily derived from unicity.


$$
\begin{array}{ll}
\begin{array}{rl}
& f = \f{foldr}\ g\ e \\
\stepe{unicity-foldr}
& \left \{
\begin{array}{l}
f\ [\ ]  =  e\\
\mbox{for all } x, xs,\\
\ \ f\ (x:xs) = g\ x\ (f\ xs)
\end{array}
\right .
\end{array}
& 
\begin{array}{rl}
  &  h \after \f{foldr}\ g\ e = \f{foldr}\ g'\ e'\\
  \stepif{foldr-fusion}
  & \left \{ 
    \begin{array}{l}
      h\ e = e'\\
      h\ \after (g\ x) = (g'\ x) \after h
    \end{array}
  \right .
\end{array}
\end{array}
$$

$$
\begin{array}{ll}
\begin{array}{rl}
& f = \f{unfoldLTree}\ g \\
\stepe{unicity-unfoldLTree}
& \mbox{for all } x,\\
&\begin{array}{l}
f\ x  =  \f{case}\ (g\ x)\ \f{of}\\
\qquad \f{Right}\ y \to \f{Leaf}\ y\\
\qquad \f{Left}\ (l,r) \to \f{Branch}\ (f\ l)\ (f\ r)
\end{array}
\end{array}
&
\begin{array}{rl}
& f = \f{unfoldBTree}\ g \\
\stepe{unicity-unfoldBTree}
& \mbox{for all } x,\\
&\begin{array}{l}
f\ x  =  \f{case}\ (g\ x)\ \f{of}\\
\qquad \f{Right}\ () \to \f{Empty}\\
\qquad \f{Left}\ (y,l,r) \to \f{Node}\ y\ (f\ l)\ (f\ r)
\end{array}
\end{array}
\end{array}
$$

\paragraph{Hylomorphisms.}

The composition of a fold over a regular type $T$ with an unfold of
that type is a recursive function whose recursion tree is shaped in
the same way as $T$. Such a definition can be
\emph{deforested}~\cite{WadlerP:dtpet}, i.e. the construction of the
intermediate data-structures can be eliminated, yielding a direct
recursive definition. As an example, the definition $h =
(\f{foldLTree}\ f\ e) \after (\f{unfoldLTree}\ g)$ can be deforested
to give:
%
\begin{code}
h x = case (g x) of
      Right y    -> e y
      Left (l,r) -> f (h l) (h r)
\end{code}

This corresponds to a new generic recursion pattern, called a
\emph{hylomorphism}. Hylomorphisms do not possess a unicity property,
but they are still useful for reasoning about programs, using the
properties of their fold and unfold components.
In particular, hylomorphisms are useful for capturing the structure of
functions that are not directly defined by structured recursion or
co-recursion, as is the case of the divide-and-conquer sorting
algorithms: the unfold component takes the unsorted list and
constructs a tree; the fold iterates over this structure to produce
the sorted list.
The sorting algorithms introduced in the previous section were studied
as hylomorhpisms in~\cite{AugusteijnL:sorm}. In the present paper we
use this hylomorphic structure to calculate these algorithms from a
common specification.



\section{Sorting by Insertion}
\label{sec:sori}

In the rest of the paper we will repeatedly apply the following
principles. Consider a type constructor $\ro{C}$ and the following
functions:
$$
\begin{array}{rcl}
  \f{istC} & : & a \to \ro{C}\ a \to \ro{C}\ a\\
  \f{C2list} & : & \ro{C}\ a \to [a]
\end{array}
$$
The idea is that $\ro{C}\ a$ is a container type for elements of type
$a$ (typically a tree-shaped type);
%
%
\tf{istC} inserts an element in a container to give a new container;
and \tf{C2list} converts a container into a sorted list of type
$a$. 

A generic sorting algorithm can then be defined, with a container
acting as intermediate data-structure. The idea is that elements are
inserted one by one by folding over the list; a sorted list is then
obtained using \tf{C2list}. $\varepsilon \dcolon \ro{C}\ a$ is an
appropriate ``empty value''.
\begin{align}
  \f{isortC} & = \f{C2list} \after (\f{foldr\ istC\
    \varepsilon})
\label{eq:sortRightLeft}
\end{align}
It is easy to see that the algorithm is correct if the intermediate
data-structure contains exactly the same elements as the initial list,
and $\f{C2list}$ somehow produces a sorted list from the elements in
the intermediate structure. This can be formalized by constructing a
proof of equivalence to insertion sort, which gives necessary
conditions for the algorithm to be correct.

$$
\begin{array}{rl}
& \f{C2list} \after (\f{foldr\ istC}\ \varepsilon) = \f{foldr\ insert}\ [\
  ]\\
\stepe{unicity-foldr}
& \left \{
\begin{array}{l}
\f{C2list}\ (\f{foldr\ istC}\ \varepsilon\ [\ ]) = [\ ]\\
\f{C2list}\ (\f{foldr\ istC\ \varepsilon}\ (x:xs)) = \f{insert}\ x\
(\f{C2list}\ (\f{foldr\ istC}\ \varepsilon\ xs))
\end{array}
\right .\\
\stepe{def. foldr}
& \left \{
\begin{array}{l}
\f{C2list}\ \varepsilon = [\ ]\\
\f{C2list}\ (\f{istC}\ x\ (\f{foldr\ istC}\ \varepsilon\ xs)) = \f{insert}\ x\
(\f{C2list}\ (\f{foldr\ istC}\ \varepsilon\ xs))
\end{array}
\right .
\end{array}
$$

Alternatively one can use fusion, which leads to stronger conditions:

$$
\begin{array}{rl}
& \f{C2list} \after (\f{foldr\ istC}\ \varepsilon) = \f{foldr\ insert}\ [\
  ]\\
\stepif{foldr fusion}
& \left \{
\begin{array}{l}
  \f{C2list}\ \varepsilon = [\ ]\\
  \f{C2list}\ \after (\f{istC}\ x) = (\f{insert}\ x) \after 
  \f{C2list}
\end{array}
\right .
\end{array}
$$

Thus for each concrete container type it is sufficient to prove
equation~\ref{eq:tl1} and one of~\ref{eq:tl2} or~\ref{eq:tl3} to
establish that the corresponding function \tf{isortC} is indeed a
sorting algorithm:
\begin{align}
\label{eq:tl1} \f{C2list}\ \varepsilon & = [\ ]\\
\label{eq:tl2} \f{C2list} \after (\f{istC}\ x) & = (\f{insert}\ x)
\after \f{C2list} \\
\label{eq:tl3} \f{C2list}\ (\f{istC}\ x\ (\f{foldr\ istC}\ \varepsilon\
xs)) & = \f{insert}\ x\ (\f{C2list}\ (\f{foldr\ istC}\ \varepsilon\ xs))
\end{align}
Note that together, equations~\ref{eq:tl1} and~\ref{eq:tl2} mean that
\tf{C2list} is a homomorphism between the structures $(C\ a, \f{istC},
\varepsilon)$ and $([a],\f{insert},[\ ])$.






Observe that the above algorithm constructs the intermediate structure
by inserting the elements from right to left.  
%
%
A tail-recursive version of \tf{isortC} can be derived by a standard
transformation based on fusion~\cite{BirdR:proastp}. This will
construct the intermediate structure in a rightwards fashion.
We start by writing a specification for this function $\f{isortC}_t$.

$$
\begin{array}{rcl}
 \f{isortC}_t & : & [a] \to \ro{C}\ a \to [a]\\
 \f{isortC}_t\ l\ y & = & (\f{isort}\ l) \odot (\f{C2list}\ y)
\end{array}
$$
The tail-recursive function uses an extra accumulator argument of the
chosen container type. In the call $\f{isortC}_t\ l\ y$, $l$ is the
list that remains to be sorted, and the accumulator $y$ contains
elements already inserted in the container. The right-hand side of the
equality states how the final result can be obtained using insertion
sort and the conversion of $y$ to a list. 

The following definition satisfies the specification (proof is given
in Appendix~\ref{app:sec:isortCt}).
$$
\begin{array}{l}
  \f{isortC}_t =  \f{foldr}\ \f{istC}'\ \f{C2list}\\
  \qquad \qquad  \f{where}\  \f{istC}'\ x\ f\ y = f\ (\f{istC}\ x\ y)\\
\end{array}
$$
Then $\f{isortC}_t\ l\ \varepsilon  = \f{iSort}\ l$ holds as an immediate
consequence of the specification and eq.~(\ref{eq:tl1}) above.
%
%
An alternative version of this can be defined, which separates the
tail-recursive construction of the intermediate structure from its
conversion to a sorted list (note $\f{ap} \varepsilon\ f = f\ \varepsilon$):
\begin{align}
%
  \f{isortC}' = \f{C2list} \after (\f{ap} \varepsilon) \after (\f{foldr}\ \f{istC}'\ \f{id})
\label{eq:sortLeftRight}
\end{align}
It is straightforward to establish that $\f{isortC}'\ l =
\f{iSortC}_t\ l\ \varepsilon$, thus $\f{isortC}' =  \f{iSort}$.


In the next sections, the container type and its empty value, together
with the functions \tf{istC} and \tf{C2list}, will be instantiated to
produce three different insertion-based algorithms, using
schemes~\ref{eq:sortRightLeft} and~\ref{eq:sortLeftRight}. 

Each algorithm will be proved correct by calculating
eqs.~\ref{eq:tl1}, and~\ref{eq:tl2} or~\ref{eq:tl3} above. The next
step will be to transform each algorithm into a hylomorphism that can
then be deforested, resulting in a well-known sorting algorithm.  For
this, it will suffice to transform the function that constructs the
intermediate tree into co-recursive form.



\section{A Derivation of Merge Sort}
\label{sec:mergesort}

Our first concrete sorting algorithm based on insertion into an
intermediate structure uses \emph{leaf-labelled} binary trees.  This
is given in Table~\ref{tab:sortinsertLT}. We remark that to cover the
case of the empty list, a Maybe type is used in the leaves of the
trees.

\begin{table}[tb]
\hrule
$$
\begin{array}{l}\\
\begin{array}{lcl}
\f{isortLT}  & = & \f{LT2list} \after \f{buildLT}\\ 
\f{buildLT} & = & \f{foldr\ istLT\ (Leaf\ Nothing)}
\end{array}
\\ \\
\begin{array}{lcl}
\f{LT2list} & = & \f{foldLTree}\ (\odot)\ t\\
&& \mbox{ where } 
\begin{array}[t]{l}
t\ \f{Nothing} = [\ ]\\
t\ (\f{Just}\ x) = [x]
\end{array}\\
\end{array}
\\ \\
\begin{array}{lcl}
  \f{istLT}\ x\ (\f{Leaf\ Nothing}) & = &  \f{Leaf}\ (\f{Just}\ x)\\
  \f{istLT}\ x\ (\f{Leaf} (\f{Just}\ y)) & = &  \f{Branch}\ (\f{Leaf}\
  (\f{Just}\ x))\ (\f{Leaf} (\f{Just}\ y))\\
  \f{istLT}\ x\ (\f{Branch}\ l\  r) & = & \f{Branch}\ (\f{istLT}\ x\ r)\
  l
\end{array}
\end{array}
$$
\hrule
\caption{Sorting by insertion in a leaf tree}
\label{tab:sortinsertLT}
\end{table}

\begin{proposition} \label{prop:P1}
$\f{isortLT}$ is a sorting algorithm.
\end{proposition}

\proof
We instantiate eqs.~(\ref{eq:tl1}) and~(\ref{eq:tl2}). Note that
the empty value here is $\varepsilon = \f{Leaf\ Nothing}$.

\begin{align*}
\f{LT2list}\ (\f{Leaf\ Nothing}) & = [\ ]\\
\f{LT2list} \after (\f{istLT}\ x) & = (\f{insert}\ x)
  \after \f{LT2list} \\
\end{align*}

The first equality if true by definition; the second can be proved by
induction, or alternatively using fusion. The latter proof is given in
Appendix~\ref{app:sec:istLT}. Together these equations establish that
$\f{LT2list}$ is a homomorphism between the structures $(\LTree a,
\f{istLT}, \f{Leaf\ Nothing})$ and $([a],\f{insert},[\ ])$.
\endproof

It is also easy to see that the intermediate tree is \emph{balanced}:
the difference between the heights of the subtrees of a node is never
greater than one, since subtrees are swapped at each insertion
step. Note that the insertion function $\f{istLT}$ was carefully
designed with efficiency in mind, which grants execution in time $O(N
\lg N)$; other solutions would still lead to sorting algorithms,
albeit less efficient.

\begin{proposition}
  The trees constructed by $\f{buildLT}$ are balanced.
\end{proposition}

\proof 
It can be proved by induction on the structure of the argument list
that either the subtrees of the constructed tree have the same height,
or the height of the left subtree is greater than the height of the
right subtree by one unit. The function $\f{istLT}$ preserves this
invariant.
\endproof

The next transformation step applies to the function that constructs
the intermediate tree.  An alternative way of constructing a balanced
tree is by \emph{unfolding}: the initial list is traversed and its
elements placed alternately in two subsequences, which are then used
as arguments to recursively construct the subtrees. Note that the
sequences will have approximately the same length. For singular and
empty lists, leaves are returned.
$$
\begin{array}{rcl}
  \f{unfoldmsort} & = & \f{unfoldLTree\ g}\\
  && \f{where} 
             \begin{array}[t]{l}
             \f{g}\ [\ ]  = \f{Right\ Nothing}\\
             \f{g}\ [x] = \f{Right}\ (\f{Just}\ x)\\
             \f{g}\ xs  = \f{Left}\ (\f{maux}\ xs)\\
             \f{maux}\ [\ ]     = ([\ ],[\ ])\\
             \f{maux}\ (h:t) = (h:b,a)\  \f{where}\ (a,b) = \f{maux}\ t\\
           \end{array}
         \end{array}
$$

\begin{proposition} The above function constructs the same
  intermediate trees as those obtained by folding over the argument
  list:
  $$
  \f{foldr\ istLT\ (Leaf\ Nothing)} = \f{unfoldmsort}
  $$
\end{proposition}

\proof

We use the unicity property of leaf-tree unfolds:

$$
\begin{array}{rl}
& \f{foldr}\ \f{istLT}\ (\f{Leaf\ Nothing}) =  \f{unfoldLTree}\ g\\  
\stepe{unicity-unfoldLTree}
& \mbox{for all } x,\\
&\begin{array}{l}
(\f{foldr}\ \f{istLT}\ (\f{Leaf\ Nothing}))\ x  =  \f{case}\ (\f{g}\
x)\ \f{of}\\  
\qquad \f{Right}\ y \to \f{Leaf}\ y\\
\qquad \f{Left}\ (l,r) \to \f{Branch}\ (\f{f}\ l)\ (\f{f}\ r)
\end{array}\\
\stepe{by cases}
& \left \{
\begin{array}{lr}
\f{Leaf\ Nothing} = \f{Leaf\ Nothing} & \mbox{ if }x = [\ ]\\
\f{Leaf} (\f{Just}\ h) = \f{Leaf} (\f{Just}\ h) & \mbox{ if } x =
[h]\\
\f{istLT}\ h_1\ (\f{foldr\ istLT\ (Leaf\ Nothing)}\ (h_2:t))\\
 = \f{Branch}\
(\f{foldr\ istLT\ (Leaf\ Nothing)}\ l) 
\\
\qquad \qquad \ (\f{foldr\ istLT\ (Leaf\ Nothing)}\ r) &  \mbox{ if }
x = h_1:h_2:t \\
\qquad \mbox{ where } (l,r) = \f{maux}\ (h_1:h_2:t)
\end{array}
\right .
\end{array}
$$
And the last equality can be easily proved by induction on the
structure of $t$.
\endproof

Substituting this in the definition of $\f{isortLT}$ yields a
hylomorphism that is of course still equivalent to insertion sort. It
is immediate to see that this can be deforested, and the result is
merge sort:
$$
\f{LT2list} \after \f{unfoldmsort} = \f{msort}
$$

\section{A Derivation of Heapsort}
\label{sec:deriv-heap-sort}

\def\foldr{\f{foldr}}
\def\foldl{\f{foldl}}
\def\inL{\f{in_L}}
\def\inT{\f{in_T}}
\def\id{\f{id}}
\def\isortBST{\f{isortBST}}
\def\BST2list{\f{BST2list}}
\def\buildBST{\f{buildBST}}
\def\buildBSTacc{\f{bAcc}}
\def\istBST{\f{istBST}}
\def\Left{\f{Left}}
\def\Right{\f{Right}}
\def\Empty{\f{Empty}}
\def\Node{\f{Node}}
\def\qaux{\f{qaux}}
\def\insert{\f{insert}}
\def\Nil{[\,]}
\def\isortH{\f{isortH}}
\def\H2list{\f{H2list}}
\def\buildH{\f{buildH}}
\def\istH{\f{istH}}
\def\haux{\f{haux}}

\begin{table}[tb]
\hrule
\[
\begin{array}{l}\\
\begin{array}{lcl}
\isortH  & = & \H2list \after \buildH\\ 
\buildH & = & \foldr\ \istH\ \Empty\\ \\
\H2list & = & \foldr\ \f{aux}\ \Nil \\ 
&& \mbox{where } \f{aux}\ x\ l\ r = x:(l \odot r)\\
\end{array} \\ \\
\begin{array}{lcl}
\istH\ x\ \Empty & = &  \Node\ x\ \Empty\ \Empty\\
\istH\ x\ (\Node\ y\ l\ r) & | & x<y\quad=\quad \Node\ x\
(\istH\ y\ r)\ l\\
  & | & \f{otherwise}\quad=\quad \Node\ y\ (\istBST\ x\ r)\ l \\
\end{array}
\end{array}
\]
\hrule
\caption{Sorting by insertion in a heap}
\label{tab:sortiH}
\end{table}

In the \emph{heapsort} algorithm, one computes the minimum of the list
prior to the recursive calls. This will determine that each node of
the intermediate structure (the recursion tree) this minimum for some
tree; it is thus a binary \emph{node-labelled} tree.

We repeat the program taken for the derivation of the merge sort: we
design a function that inserts a single element in the intermediate
tree ($\istH$), iterate this function over the initial list
($\buildH$) and then provide a function that recovers the ordered list
from the tree ($\H2list$). These functions are shown in Table
\ref{tab:sortiH}.

\begin{proposition}
  $\isortH$ is a sorting algorithm.  
\end{proposition}
\begin{proof}
  We instantiate eqs.~(\ref{eq:tl1}) and~(\ref{eq:tl3}). We set
  $\epsilon = \Empty$, and thus eq.~(\ref{eq:tl1}) results directly
  from the definition. For eq.~(\ref{eq:tl3}), we need to prove that
  for every list $l$,
  \[ \insert\ x\ (\H2list\ (\buildH\ l)) = \H2list\ (\istH\ x \
  (\buildH\ l)). \] In order to prove this, we rely on the fact that
  trees generated by $\buildH$ are always \emph{heaps}, i.e. the root
  element is the least of the tree. The complete derivation is
  presented in appendix \ref{app:sec:tree-invs} (Propositions
  \ref{prop:h2ist} and \ref{prop:buildh}).
\end{proof}  

Note that in order to prove the correctness of this algorithm, we
cannot rely on the strongest hypothesis given by eq. \ref{eq:tl2}
(obtained from the use of the fusion law) as we have done for merge
sort. The reason for this is that, for an arbitrary tree $t$,
\[ \insert\ x\ (\H2list\ t) \not= \H2list\ (\istH\ x\ t). \]
\noindent On the other hand, the weaker requisite given by
eq. \ref{eq:tl3} (obtained by the use of unicity or induction) retains
the information that we restrict our attention to trees constructed by
$\buildBST$, and these will satisfy the required equality.

We also note that the intermediate tree is again balanced (essentially
by the same argument used for merge sort). This means that this
sorting algorithm also executes in time $O(N\lg N)$.

It remains to show that the intermediate tree can be constructed
coinductively. For that, consider the following function:

\[
\begin{array}{lcl}
  \f{unfoldhsort} & = & \f{unfoldBTree\ g}\\
  &&\f{where}
  \begin{array}[t]{lcl}
    \f{g}\ \Nil  & = & \Right\ () \\
    \f{g}\ (x:xs)  & = & \Left\ (\haux\ x\ xs)\\
    \haux\ x\ \Nil  & = & (x,\Nil,\Nil) \\
    \haux\ x\ (y:ys) & | & x<m = (x,m:b,a) \\
    & | & \f{otherwise} = (m,x:b,a)\\
    & & \qquad \f{where}\ (m,a,b) = \haux\ y\ ys
  \end{array} 
\end{array}
\]


\begin{proposition}
  \label{prop:buildH-unfold}
  The above function constructs the same intermediate trees as those
  obtained by folding over the argument list:
  $$
  \buildH = \f{unfoldhsort}
  $$
\end{proposition}
\begin{proof}
\[
  \begin{array}{rl}
    &\f{unfoldhsort} = \f{buildH}\\
    \stepe{by $\f{unicity-unfoldBTree}$}
    &
    \left \{
      \begin{array}[c]{l}
        \buildH\ \Nil = \Empty \\
        \buildH\ (x:xs) 
        = \Node\ z\ (\buildH\ a)\ (\buildH\ b) \\
        \quad\mbox{ where } (z,a,b) = \haux\ x\ xs
      \end{array}
    \right . \\
    \stepe{definitions}
    &
    \left \{
      \begin{array}[c]{l}
        \Empty = \Empty \\
        \istH\ x\ (\buildH\ xs) 
        = \Node\ z\ (\buildH\ a)\ (\buildH\ b) \\
        \quad\mbox{ where } (z,a,b) = \haux\ x\ xs
      \end{array}
    \right . 
  \end{array}
\]

The second equality is proved by structural induction on $xs$.
For the base case ($xs=\Nil$), it follows directly from evaluating the
definitions. For the inductive step ($xs=y:ys$),
let us assume that $(z,a,b)=(\haux\ y\ ys)$. The definition of $\haux$
tell us that
\[
(z',a',b') = (\haux\ x\ (y:ys)) =
\begin{cases}
(x,z:b,a) &\text{if $x<z$}, \\
(z,x:b,a) &\text{if $x\geq z$}.  
\end{cases}
\]
Thus,
\[
  \begin{array}{rl}
    & \istH\ x\ (\buildH\ (y:ys)) \\
    \stepi{def. $\buildH$}
    & \istH\ x\ (\istH\ y\ (\buildH\ ys)) \\
    \stepi{induction hypotheses}
    & \istH\ x\ (\Node\ z\ (\buildH\ a)\ (\buildH\ b)) \\
    \stepi{def. $\istH$}
    & \begin{cases}
      \Node\ x\ (\istH\ z\ (\buildH\ b))\ (\buildH\ a)
      &\text{if $x<z$}, \\
      \Node\ z\ (\istH\ x\ (\buildH\ b))\ (\buildH\ a)
      &\text{if $x\geq z$}. \\
      \end{cases}\\
    \stepi{def. of $z',a',b'$}
    &\Node\ z'\ (\buildH\ a')\ (\buildH\ b')
 \end{array}
\]
\end{proof}

As would be expected, the hylomorphism obtained replacing $\buildH$ by
$\f{unfoldhsort}$ can be deforested, and the result is the original
$\hsort$.
\[ \H2list\after\f{unfoldhsort} = \f{hsort} \]


\section{A Derivation of Quicksort}
\label{sec:deriv-quick-sort}

\begin{table}[tb]
\hrule
\[
\begin{array}{l}\\
\begin{array}{lcl}
\isortBST  & = & \BST2list \after \buildBST\\
\buildBST & = & (\ap\ \Empty) \after \buildBSTacc\\
&& \mbox{ where } 
\begin{array}[t]{l}
\ap\ x\ f = f\ x  \\
\buildBSTacc = \foldr\ \f{aux}\ \id \\
\f{aux}\ x\ f\ a = f\ (\istBST\ x\ a)
\end{array}
\end{array}
\\ \\
\begin{array}{lcl}
\BST2list & = & \f{foldBTree}\ \f{aux}\ \Nil\\
&& \mbox{ where } \f{aux}\ x\ l\ r = l \app (x:r)
\end{array}
\\ \\
\begin{array}{lcl}
  \istBST\ x\ (\Empty) & = &  \Node\ x\ \Empty\ \Empty\\
  \istBST\ x\ (\Node\ y\ l\ r) & | & x<y\quad=\quad \Node\ y\
  (\istBST\ x\ l)\ r\\
  & | & \f{otherwise}\quad=\quad \Node\ y\ l\ (\istBST\ x\ r)
\end{array}
\end{array}
\]
\hrule
\caption{Sorting by insertion in a binary search tree}
\label{tab:sortiBST}
\end{table}

In the quicksort algorithm, the activity performed prior to the
recursive calls is different from that in heapsort: instead of finding
the minimum of the list, the head of the list is used as a
\emph{pivot} for splitting the tail. Again, the intermediate structure
is a \emph{node-labelled} binary tree. But now, its ordering
properties are different --- the constructed trees will be
\emph{binary search trees}, and it suffices to traverse these trees
\emph{in-order} to produce the desired sorted list.

Following the same line as in the derivation of the previous
algorithms, we define an algorithm that iteratively inserts elements
from a list into a binary tree and then reconstructs the list by the
\emph{in-order} traversal.  This algorithm is given in Table
\ref{tab:sortiBST}.

Observe that this algorithm iterates on the initial list from left to
right (we may think of it as using the Haskell $\foldl$ operator, but
we write it as a higher-order function using $\foldr$, to exploit the
application of the rules presented earlier). This will become evident
below when we replace this function by one that constructs the
intermediate tree corecursively. For the correctness argument, we know
that the order of traversal for the initial list is irrelevant (as
shown in Section \ref{sec:sori}).

\begin{proposition}
  $\isortBST$ is a sorting algorithm.  
\end{proposition}
\begin{proof}
  We instantiate eqs. \ref{eq:tl1} and \ref{eq:tl3}. We set $\epsilon
  = \Empty$, and thus Equation \ref{eq:tl1} results directly from the
  definition. For Equation \ref{eq:tl3}, we need to prove that for
  every list,
  \[ \insert\ x\ (\BST2list\ (\buildBST\ l)) = \BST2list\ (\istBST\ x
  \ (\buildBST\ l)). \] In order to prove this, we rely on the fact
  that trees generated by $\buildBST$ are always binary search
  trees. The complete derivation is presented in appendix
  \ref{app:sec:tree-invs} (Propositions \ref{prop:bst2ist} and
  \ref{prop:buildbst}).
\end{proof}

To obtain the well-known quicksort algorithm, we need to replace the
iterated insertion function $\buildBST$ by an unfold.
\[
\begin{array}{lcl}
  \f{unfoldqsort} & = & \f{unfoldBTree\ g}\\
  && \f{where} 
             \begin{array}[t]{lcl}
               \f{g}\ \Nil  & = & \Right\ () \\
               \f{g}\ (x:xs) & = &\Left\ (\qaux\ x\ xs)\\
               \qaux\ x\ \Nil  & = & (x,\Nil,\Nil) \\
               \qaux\ x\ (y:ys)  & | & y<x = (x,y:b,a) \\
               & | & \f{otherwise} = (m,a,y:b) \\
               && \qquad\f{where}\ (a,b) = \qaux\ x\ ys
             \end{array} \\
\end{array}
\]

\begin{proposition}
\label{prop:buildBST-unfold}
  The above function constructs the same intermediate trees as those
  obtained by folding over the argument list:
  $$
  \buildBST = \f{unfoldqsort}
  $$
\end{proposition}
\begin{proof}
\[
  \begin{array}{rl}
    &\f{unfoldqsort} = \f{buildBST}\\
    \stepe{unicity-unfolBTree}
    &
    \left \{
      \begin{array}[c]{l}
        \buildBST\ \Nil = \Empty \\
        \buildBST\ (x:xs) 
        = \Node\ x\ (\buildBSTacc\ a\ \Empty)\ (\buildBSTacc\ b\
        \Empty) \\
        \quad\mbox{ where } (a,b) = \qaux\ x\ xs
      \end{array}
    \right .
    \\
    \stepe{definitions}
    &
    \left \{
      \begin{array}[c]{l}
        \buildBSTacc\ \Nil\ \Empty = \Empty \\
        \buildBSTacc\ (x:xs)\ \Empty
        = \Node\ x\ (\buildBSTacc\ a\ \Empty)\ (\buildBSTacc\ b\
        \Empty) \\
        \quad\mbox{ where } (a,b) = \qaux\ x\ xs
      \end{array}
    \right . \\
    \stepe{simplification}
    &
    \left \{
      \begin{array}[c]{l}
        \Empty = \Empty \\
        \buildBSTacc\ xs\ (\Node\ x\ \Empty\ \Empty)
        = \Node\ x\ (\buildBSTacc\ a\ \Empty)\ (\buildBSTacc\ b\
        \Empty) \\
        \quad\mbox{ where } (a,b) = \qaux\ x\ xs
      \end{array}
    \right .
  \end{array}
\]

We prove the second equality in a slightly strengthened formulation.
  For every tree $(\Node\ x\ l\ r)$ and list $xs$,
\begin{gather*}
 \buildBSTacc\ xs\ (\Node\ x\ l\ r)
= \Node\ x\ (\buildBSTacc\ a\ l)\ (\buildBSTacc\ b\ r) \\
\qquad\qquad\text{where } (a,b) = \qaux\ x\ xs
\end{gather*}
By induction on the structure of $xs$. For the base case ($xs=\Nil$),
it follows directly from evaluating the definitions. For the inductive
step ($xs=y:ys$), we reason by cases. If $x<y$, then
\[
  \begin{array}{rl}
    & \buildBSTacc\ (y:ys)\ (\Node\ x\ l'\ r') =
    \Node\ x\ (\buildBSTacc\ a'\ l')\ (\buildBSTacc\ b'\ r') \\
    &\quad\mbox{ where } (a',b') = \qaux\ x\ (y:ys) \\
    \stepe{definition, $x<y$}
    & \buildBSTacc\ ys\ (\istBST\ y\ (\Node\ x\ l'\ r') =
    \Node\ x\ (\buildBSTacc\ a\ l')\ (\buildBSTacc\ (y:b)\ r') \\
    &\quad\mbox{ where } (a,b) = \qaux\ x\ ys \\
    \stepe{definition, $x<y$}
    & \buildBSTacc\ ys\ (\Node\ x\ l'\ (\istBST\ y\ r') =
    \Node\ x\ (\buildBSTacc\ a\ l')\ (\buildBSTacc\ b\ (\istBST\ y\ r')) \\
    &\quad\mbox{ where } (a,b) = \qaux\ x\ ys \\
    \stepe{induction hypoptheses}
    & \buildBSTacc\ ys\ (\Node\ x\ l'\ (\istBST\ y\ r') =
    \buildBSTacc\ ys\ (\Node\ x\ l'\ (\istBST\ y\ r')) \\
    &\quad\mbox{ where } (a,b) = \qaux\ x\ ys \\
  \end{array}
\]
Similarly for the case ($x\geq y$).  This concludes the proof.

\end{proof}

We conclude with the statement that the hylomorphism obtained is, as
intended, the forested version of the original quicksort algorithm.
\[ \BST2list\after\f{unfoldqsort} = \f{qsort} \]

\section{Conclusion}
\label{sec:con}



This paper illustrates the strengths of the ``program calculation''
style of reasoning, in particular the simplicity of using the unicity
property of unfolds as an alternative to using coinductive principles
based on bissimulations, and more generally the structural aspects of
proofs. Inductive proofs are however often much simpler to carry out
than using the equational style, so we are not dogmatic about the
style in which proofs are presented.

Apart from the proofs of correctness which as far as we know are new,
the contributions of this paper include (two versions of) a
\emph{generic} sorting algorithm, of which 3 concretizations are used.
The role played by structural invariants in this study should also be
emphasized. 

Even when they are not crucial to the calculations, invariants provide
a much more natural setting for conducting them. Morevoer,
\emph{efficiency} properties of the algorithms, which we have left out
of this study, can only be established using \emph{well-balancing}
invariants on the intermediate trees (these invariants can easily be
proved by induction for both \tf{isortLT} and \tf{isortH}, which run
in time $O(N\lg N)$).

Another application of invariants would come up in a \emph{generic
  programming} setting: the \tf{C2list} functions would have a single
definition for every tree type: the function would merge together the
lists resulting from recursive calls with the (wrapped) contents of
nodes and leaves. For each concrete intermediate type, the structural
invariants would then allow us to refine the definition into the one
given in this paper.

This study opens the way to a richer interplay between invariants and
recursion patterns -- a topic that is not explored in this paper, but
is being currently investigated by the authors.

Finally, we have left completely out of the paper a study of
\emph{stability} of the sorting algorithms, an important property in
the presence of data-types for which the order is not total. Some of
the algorithms derived are stable and others are not, which means that
under this premise, which invalidates commutativity of $\odot$, they
are not all equivalent.

\nocite{MeertensL:par}
\nocite{MeijerE:funpbleb}

\bibliography{bibfile}

\newpage

\appendix

\section{Proofs and Calculations}

\subsection{}
\label{app:sec:isortCt}

The function
$$
\begin{array}{l}
  \f{isortC}_t =  \f{foldr}\ \f{istC}'\ \f{C2list}\\
  \qquad \qquad  \f{where}\   \f{istC}'\ x\ f\ y = f\ (\f{istC}\ x\ y)
\end{array}
$$
satisfies the specification
$$
\begin{array}{rcl}
 \f{isortC}_t & : & [a] \to \ro{C}\ a \to [a]\\
 \f{isortC}_t\ l\ y & = & (\f{isort}\ l) \odot (\f{C2list}\ y)
\end{array}
$$

\proof
The specification can be rewritten as
$$
\f{isortC}_t\ l\ y = (\f{isort}\ l) \oplus y
$$
or 
$$
\f{isortC}_t = (\oplus) \after \f{isort}
$$
with the $\oplus$ operator defined as
$$
s \oplus y= s \odot (\f{C2list}\ y)
$$
This appeals to the use of the fusion law since \tf{isort} is defined
as a fold.
$$
\begin{array}{rl}
  & \f{isortC}_t = (\oplus) \after \f{isort}\\
  \stepe{definitions}
  & \f{foldr}\ \f{istC}'\ \f{C2list} = (\oplus) \after (\f{foldr}\ \f{insert}\ [\ ])\\
  \stepif{\tf{foldr} fusion, with $\oplus$ strict}
  & \left \{ 
    \begin{array}{l}
    (\oplus)\ [\ ] = \f{C2list}\\
    (\oplus) \after (\f{insert}\ x) = (\f{istC}'\ x) \after (\oplus)
  \end{array}
  \right .\\
  \stepe{$\eta$-expansion}
  & \left \{ 
    \begin{array}{l}
  [\ ] \oplus y = \f{C2list}\ y\\
  (\f{insert}\ x\ l) \oplus y = \f{istC}'\ x\ ((\oplus)\ l)\ y
  \end{array}
  \right .\\
  \stepe{def. $\oplus$, properties of $\odot$, def. \tf{istC}'} 
  & \left \{ 
    \begin{array}{l}
  \f{C2list}\ y = \f{C2list}\ y\\
  (\f{insert}\ x\ l) \oplus y = l \oplus (\f{istC}\ x\ y)
  \end{array}
  \right .\\
  \stepe{eq.\eqref{eq:inswrap1}, def. $\oplus$}
  & [x] \odot l \odot (\f{C2list}\ y) = l \odot (\f{C2list}\ (\f{istC}\ x\ y))\\
  \stepe{eq.~(\ref{eq:tl2})}
  & [x] \odot l \odot (\f{C2list}\ y) = l \odot (\f{insert}\ x\
  (\f{C2list}\ y))\\
  \stepe{eq.\eqref{eq:inswrap1}, properties of $\odot$}
  & [x] \odot l \odot (\f{C2list}\ y) = [x] \odot l \odot (\f{C2list}\ y)
\end{array}
$$
\endproof

\subsection{}
\label{app:sec:istLT}



We prove
$$
\f{LT2list} \after (\f{istLT}\ x)  = (\f{insert}\ x) \after
\f{LT2list}
$$

first by calculation, and then using induction. 

\paragraph{Proof by Calculation.}

It is easy to see that the insertion function \tf{istLT} cannot be
written as a fold over trees, since it uses one of the subtrees
unchanged (insertion will proceed recursively in the other
subtree). This is a typical example of a situation where iteration is
not sufficient: primitive recursion is required. This has been studied
as the \emph{paramorphism} recursion pattern~\cite{MeertensL:par}. The
operator in Table~\ref{tab:list-para} embodies this pattern for
leaf-trees. The corresponding unicity property and fusion
law~\cite{MeijerE:funpbleb} are also shown in the table.

\begin{table}[tb]
\hrule
\begin{code}
paraLTree :: ((LTree a)->b->(LTree a)->b->b)-> ((Maybe a)->b)-> LTree a-> b
paraLTree f g (Leaf x)     = g x
paraLTree f g (Branch l r) = f l (paraLTree f g l) r (paraLTree f g r)
\end{code}
\hrule

$$
\begin{array}{c|c}
\begin{array}[t]{rl}
& h = \f{paraLTree}\ f\  g\\
\stepe{unicity-paraLTree}
& \left \{
\begin{array}{l}
h \after \f{Leaf}  =  g \\
\mbox{for all } l, r,\\
\ \ h\ (\f{Branch}\ l\ r) = f\ l\ (h\ l)\ r\ (h\ r)
\end{array}
\right .
\end{array}
$$
&
$$
\begin{array}[t]{rl}
& h \after (\f{paraLTree}\ f\ g) = \f{paraLTree}\ a\ b\\
\stepe{paraLTree-fusion}
& h \mbox{ strict}\ \wedge\  h \after g = b\ \wedge\\ 
& h(f\ l\ l'\ r\ r') = a\ l\ (h\ l')\ r\ (h\ r')  
\end{array}
\end{array}
$$
\hrule
\caption{The list paramorphism recursion pattern and laws}
\label{tab:list-para}
\end{table}

The function $\f{istLT}\ x$ can now be written as the following
paramorphism of leaf trees
$$
\begin{array}{rcl}
  \f{istLT}\ x & = &\f{paraLTree}\ f\ g\\
  && \mbox{where} 
  \begin{array}[t]{l}
    g\ \f{Nothing} = \f{Leaf}(\f{Just}\ x)\\
    g\ (\f{Just}\ y) = \f{Branch}\ (\f{Leaf}(\f{Just}\ x))\
    (\f{Leaf}(\f{Just}\ y)) \\   
    f\ l\ l'\ r\ r'\ = \f{Branch}\ r'\ l\\
\end{array}\\
\end{array}
$$

We use the following strategy: we apply fusion to prove the left-hand
side of the equality equivalent to a new paramorphism; subsequently we
prove by unicity that the right-hand side of the equality is also
equivalent to this paramorphism.
$$
  \begin{array}{rl}
    & \f{LT2ist} \after (\f{istLT}\ x) = \f{paraLTree}\ a\ b\\
    \stepe{def. of $\f{istLT}\ x$ as a paramorphism}
    & \f{LT2list} \after (\f{paraLTree}\ f\ g)  = \f{paraLTree}\ a\
    b \\
    \stepif{paraLTree-fusion, with \tf{LT2list}
      strict} 
    & \left \{ 
      \begin{array}{l}
        \f{LT2list} \after g = b \\
        \f{LT2list}(f\ l\ l'\ r\ r') = a\ l\ (\f{LT2list}\ l')\ r\
        (\f{LT2list}\ r')
      \end{array}
      \right .\\
      \stepe{$\eta$-expansion, def. $f,g$}
    & \left \{ 
      \begin{array}{l}
        \f{LT2list}\ (\f{Leaf}(\f{Just}\ x)) = b\ \f{Nothing}\\
        \f{LT2list}\ (\f{Branch}\ (\f{Leaf}(\f{Just}\
        x))\ (\f{Leaf}(\f{Just}\ y))) = b\ (\f{Just}\ y)\\
        \f{LT2list}\ (\f{Branch}\ r'\ l) = a\ l\
        (\f{LT2list}\ l')\ r\ (\f{LT2list}\ r')
      \end{array}
      \right .\\
    \stepe{def. \tf{LT2list}}
    & \left \{
    \begin{array}{l}
      [x] = b\ \f{Nothing} \\
      \protect[x]  \odot [y] = b\ (\f{Just}\ y)\\
      (\f{LT2list}\ r') \odot (\f{LT2list}\ l) = a\ l\
      (\f{LT2list}\ l')\ r\ (\f{LT2list}\ r')
    \end{array}
  \right .\\
\end{array}
$$

We are thus led to define
$$
  \begin{array}{rl}
    & b\ \f{Nothing} = [x]\\
    & b\ (\f{Just}\ y) = [x] \odot [y]\\
    & a\ l\ l''\ r\ r'' = r'' \odot (\f{LT2list}\ l)
  \end{array}
$$

It remains to prove $(\f{insert}\ x) \after \f{LT2list} =
\f{paraLTree}\ a\ b$. Again we proceed by using fusion; the trick is
now to write the fold \tf{LT2list} as a paramorphism (this is always
possible since it is a particular case).
$$
  \begin{array}{rl}
    &(\f{insert}\ x) \after \f{LT2list} =
    \f{paraLTree}\ a\ b\\
    \stepe{unicity-paraLTree, with
      $(\f{insert}\ x)$  strict} 
    & 
    \left \{
      \begin{array}{l}
        (\f{insert}\ x) \after \f{LT2list} \after \f{Leaf} = b\\
        (\f{insert}\ x)\ (\f{LT2list}\ (\f{Branch}\ l\ r))  
        = a\ l\ (\f{insert}\ x\ (\f{LT2list}\ l))\ r\ (\f{insert}\ x\
        (\f{LT2list}\ r))\\ 
      \end{array}
    \right .\\
    \stepe{def. of $\f{LT2list}$}
    & 
    \left \{
      \begin{array}{l}
        (\f{insert}\ x) \after g = b\\
        \f{insert}\ x\ (f\ l\ (\f{LT2list}\ l)\ r\ (\f{LT2list}\ r))
        = a\ l\ (\f{insert}\ x\ (\f{LT2list}\ l))\ r\ (\f{insert}\ x\
        (\f{LT2list}\ r))\\ 
      \end{array}
    \right .\\
    & \mbox{ where} \\
    & \qquad g\ \f{Nothing} = [\ ]\\
    & \qquad g\ (\f{Just}\ y) = [y]\\
    & \qquad f\ l\ l'\ r\  r' = l' \odot r'\\
    \stepe{$\eta$-expansion, def. of $f,g,a,b$}
    & \left \{
      \begin{array}{l}
        \f{insert}\ x\ [\ ] = [x]\\
        \f{insert}\ x\ [y] = [x] \odot [y]\\
        \f{insert}\ x\ (\f{LT2list}\ l \odot \f{LT2list}\ r) =
        (\f{insert}\ x\ (\f{LT2list}\ r)) \odot (\f{LT2list}\ l)
      \end{array}
    \right .\\
    \stepe{\eqref{eq:inswrap1} and properties of $\odot$}
    & \left \{
      \begin{array}{l}
        \protect[x] = [x] \\
        \protect[x] \odot [y] = [x] \odot [y] \\
        \protect[x] \odot (\f{LT2list}\ l) \odot (\f{LT2list}\ r) =
        \protect[x] \odot (\f{LT2list}\ l) \odot (\f{LT2list}\ r) 
      \end{array}
    \right .
  \end{array}
$$

\paragraph{Proof by Induction.}

\begin{enumerate}
\item $c = \f{Leaf\ Nothing}$
\begin{align*}
  & \f{LT2list}\ (\f{istLT}\  x\ (\f{Leaf\ Nothing})) = \f{insert}\
  x\ (\f{LT2list}\ (\f{Leaf\ Nothing}))\\
  \stepe{def. \tf{istLT}, \tf{LT2list}}
  & \f{LT2list}\ (\f{Leaf}\ (\f{Just}\ x)) = \f{insert}\
  x\ [\ ]\\
  \stepe{def. \tf{LT2list}, \tf{insert}} 
  & [x] = [x]
\end{align*}

\item $c = \f{Leaf}\ (\f{Just}\ y)$
\begin{align*}
  & \f{LT2list}\ (\f{istLT}\  x\ (\f{Leaf}\ (\f{Just}\ y))) = \f{insert}\
  x\ (\f{LT2list}\ (\f{Leaf}\ (\f{Just}\ y)))\\
  \stepe{def. \tf{istLT}, \tf{LT2list}}
  & \f{LT2list}\ (\f{Branch}\ (\f{Leaf}\ (\f{Just}\ x))\ (\f{Leaf}\
  (\f{Just}\ y))) = \f{insert}\  x\ [y]\\
  \stepe{def. \tf{LT2list}, Spec. theorem}
  & (\f{LT2list}\ (\f{Leaf}\ (\f{Just}\ x))) \odot (\f{LT2list}\ (\f{Leaf}\
  (\f{Just}\ y))) = (\f{wrap}\ x) \odot [y]\\
  \stepe{def. \tf{LT2list}, \tf{wrap}}
  & [x] \odot [y] = [x] \odot [y]
\end{align*}

\item $c = \f{Branch}\ l\ r$
\begin{align*}
  & \f{LT2list}\ (\f{istLT}\ x\ (\f{Branch}\ l\ r)) = \f{insert}\ x\
  (\f{LT2list}\ (\f{Branch}\ l\ r))\\
  \stepe{def. \tf{istLT}, \tf{LT2list}} 
  & \f{LT2list}\ (\tf{Branch}\ (\tf{istLT}\ x\ r)\ l) =
  \f{insert}\ x\ ((\f{LT2list}\ l) \odot (\f{LT2list}\ r))\\
  \stepe{def. \tf{LT2list}, Spec. theorem} 
  & (\f{LT2list}\ (\tf{istLT}\ x\ r)) \odot (\f{LT2list}\ l) =
  (\f{wrap}\ x) \odot ((\f{LT2list}\ l) \odot (\f{LT2list}\ r))\\
  \stepe{induction, commut. $\odot$}
  & (\f{insert}\ x\ (\f{LT2list}\ r)) \odot (\f{LT2list}\ l) =
  (\f{wrap}\ x) \odot ((\f{LT2list}\ r) \odot (\f{LT2list}\ l)) \\
  \stepe{Spec. theorem, assoc. $\odot$}
   & (\f{wrap}\ x) \odot (\f{LT2list}\ r) \odot (\f{LT2list}\ l) = 
  (\f{wrap}\ x) \odot (\f{LT2list}\ r) \odot (\f{LT2list}\ l) \\
\end{align*}
\end{enumerate}

\newpage
\section{Tree Invariants}
\label{app:sec:tree-invs}

In order to prove certain equalities, it is convenient to introduce a
notion of \emph{invariant} that captures properties satisfied by the
intermediate structures. These invariants are defined structurally on
the data types.

\def\AllL{\f{AllL}}
\def\AllT{\f{AllT}}
\def\Prop{\f{Bool}}
For every predicate $p:A\rightarrow \Prop$, we consider the following
inductive predicates:
\begin{gather*}
  (\AllL\ p\ \Nil) \\
  \forall x, xs . (p\ x) \wedge (\AllL\ p\ xs) \Rightarrow
  (\AllL\ p\ (x:xs)) \\
  \\
  (\AllT\ p\ \Empty) \\
  \forall x, l, r . (p\ x) \wedge (\AllT\ p\ l) \wedge (\AllT\ p\ r)
  \Rightarrow  (\AllT\ p\ (\Node\ x\ l\ r) \\
  \\
  (\f{BST}\ \Empty) \\
  \forall x, l, r . (\AllT\ (< x)\ l) \wedge (\AllT\ (\geq x)\ r)
  \wedge (\f{BST}\ l) \wedge (\f{BST}\ r) \Rightarrow
  (\f{BST}\ (\Node\ x\ l\ r)) \\
  \\
  (\f{HEAP}\ \Empty) \\
  \forall x, l, r . (\AllT\ (\geq x)\ l) \wedge (\AllT\ (\geq x)\ r)
  \wedge (\f{HEAP}\ l) \wedge (\f{HEAP}\ r) \Rightarrow
  (\f{HEAP}\ (\Node\ x\ l\ r))
\end{gather*}

Let us start stating some simple properties concerning lists and trees.

\begin{lemma}
\label{lemma:app-props}
  For every values $x,y$ and lists $l_1.l_2$, we have:
  \begin{enumerate}
  \item $x<y \Rightarrow \insert\ x\ (l_1\app[y]\app l_2) = (\insert\
    x\ l_1)\app[y]\app l_2$
  \item $(\AllL\ (\leq\ x)\ l_1) \Rightarrow \insert\ x\ (l_1\app l_2)
    = l_1\app(\insert\ x\ l_2)$
  \item $(\forall x . p\ x \Rightarrow q\ x) \Rightarrow \AllL\ p\ l_1
    \Rightarrow \AllL\ q\ l_1$
  \item $(\AllL\ p\ (l_1\app l_2)) \Leftrightarrow (\AllL\ p\ l_1) \wedge
    (\AllL\ p\ l_2)$
  \item $(\AllL\ (<x)\ l_1) \Rightarrow \insert\ x\ l_1 = x:l_1$
  \end{enumerate}
\end{lemma}
\begin{proof}
  Simple induction on $l_1$.
\end{proof}

\begin{lemma}
\label{lemma:allT-props}
For every tree $t$ and value $x$,
  \begin{enumerate}
  \item $(\AllT\ p\ t) \Rightarrow\ (\AllL\ p\ (\BST2list\ t))$
  \item $(\AllT\ p\ t) \Rightarrow\ (\AllL\ p\ (\H2list\ t))$
  \item $(p\ x) \wedge (\AllT\ p\ t) \Rightarrow (\AllT\ p\ (\istBST\ x\
    t)$
  \item $(p\ x) \wedge (\AllT\ p\ t) \Rightarrow (\AllT\ p\ (\istH\ x\
    t)$
  \end{enumerate}
\end{lemma}
\begin{proof}
  Induction on $t$.
\end{proof}

We are now able to prove the required properties. For heapsort, we
explore the fact that the intermediate structure is a heap (its root
keeps the least element).

For the heapsort algorithm, we explore the fact that the intermediate
tree is a heap.

\begin{proposition}
\label{prop:h2ist}
  \[ (\f{HEAP}\ t) \Rightarrow \insert\ x\ (\H2list\ t) = \H2list\
  (\istH\ t) \]
\end{proposition}
\begin{proof}
  By induction on the structure of $t$. The base case follows
  immediately from the definitions. For the induction step we have:
\[
  \begin{array}{rl}
    & \insert\ x\ (\H2list\ (\Node\ y\ l\ r)) \\
    \stepi{def. $\H2list$}
    & \insert\ x\ (y:(\H2list\ l) \odot (\H2list\ r)) \\
    \stepi{def. $\insert$}
    & 
    \begin{cases}
      x:y:((\H2list\ l)\odot(\H2list\ r))
      & \text{if $x<y$,}\\
      y:(\insert\ x\ ((\H2list\ l)\odot(\H2list\ r)))
      & \text{if $x\geq y$,}\\
    \end{cases} \\
    \stepi{lemma \ref{lemma:app-props} (5)}
    & 
    \begin{cases}
      x:(\insert\ y\ ((\H2list\ l)\odot(\H2list\ r)))
      & \text{if $x<y$,}\\
      y:(\insert\ x\ ((\H2list\ l)\odot(\H2list\ r)))
      & \text{if $x\geq y$,}\\
    \end{cases} \\
    \stepi{commutativity and associativity of $\odot$}
    & 
    \begin{cases}
      x:(([y]\odot(\H2list\ r))\odot(\H2list\ l))
      & \text{if $x<y$,}\\
      y:(([x]\odot(\H2list\ r))\odot(\H2list\ l))
      & \text{if $x\geq y$,}\\
    \end{cases} \\
    \stepi{induction hypotheses}
    & 
    \begin{cases}
      x:((\H2list\ (\istH\ y\ l))\odot(\H2list\ r))
      & \text{if $x<y$,}\\
      y:((\H2list\ (\istH\ x\ l))\odot(\H2list\ r))
      & \text{if $x\geq y$,}\\
    \end{cases} \\
    \stepi{def. $\H2list$}
    &
    \begin{cases}
      \H2list\ (\Node\ x\ (\istH\ y\ l)\ r)
      & \text{if $x<y$,}\\
      \H2list\ (\Node\ y\ (\istH\ x\ l)\ r)
      & \text{if $x\geq y$,}\\
    \end{cases} \\
    \stepi{def. $\istH$}
    &
    \H2list\ (\istH\ x\ (\Node\ y\ l\ r)
 \end{array}
\]
\end{proof}

To prove that the intermediate tree is actually a heap,
we prove that insertion of elements preserves the invariant.

\begin{proposition}
For every value $x$ and tree $t$,
\[ (\f{HEAP}\ t) \Rightarrow (\f{HEAP}\ (\istH\ x\ t))). \]
\end{proposition}
\begin{proof}
  Induction on $t$. The base case follows immediately from the
  definitions. For the induction step we have:
\[
  \begin{array}{rl}
    & (\f{HEAP}\ (\istH\ x\ (\Node\ y\ l\ r))) \\
    \stepe{def. $\istH$}
    & 
    \begin{cases}
      (\f{HEAP}\ (\Node\ x\ (\istH\ y\ r)\ l))
      & \text{if $x<y$,}\\
      (\f{HEAP}\ (\Node\ y\ (\istH\ x\ r)\ l))
      & \text{if $x\geq y$,}\\
    \end{cases}
  \end{array}
\]
\noindent In fact, when $x<y$ we have:
\[
\begin{cases}
  (\AllT\ (\geq x)\ (\istH\ y\ l))
  &\text{by lemma \ref{lemma:allT-props} (2) and $(\f{HEAP}\ (\Node\ y\
    l\ r))$} \\
  (\AllT\ (\geq x)\ r)
  &\text{by $(\f{HEAP}\ (\Node\ y\ l\ r))$} \\
  (\f{HEAP}\ (\istH\ x\ r))
  &\text{by induction hypotheses and $(\f{HEAP}\ (\Node\ y\ l\ r))$} \\
  (\f{HEAP}\ l)
  &\text{by $(\f{HEAP}\ (\Node\ y\ l\ r))$}
\end{cases}
\]
\noindent We reason similarly when $x\geq y$.
\end{proof}

And now, the required result follows directly by induction.

\begin{corollary}
\label{prop:buildh}
For every list $l$,
\[ (\f{HEAP}\ (\buildH\ l)) \]
\end{corollary}
\begin{proof}
  Simple induction on $l$.
\end{proof}

For the quicksort algorithm, we explore the fact that the intermediate
tree is a binary search tree.

\begin{proposition}
\label{prop:bst2ist}
For every value $x$ and tree $t$,
  \[ (\f{BST}\ t) \Rightarrow \insert\ x\ (\BST2list\ t) = \BST2list\
  (\istBST\ t) \]
\end{proposition}
\begin{proof}
  By induction on the structure of $t$. The base case follows
  immediately from the definitions. For the induction step we have:
\[
  \begin{array}{rl}
    & \insert\ x\ (\BST2list\ (\Node\ y\ l\ r)) \\
    \stepi{def. $\BST2list$}
    & \insert\ x\ ((\BST2list\ l)\app[y]\app(\BST2list\ r)) \\
    \stepi{lemma \ref{lemma:app-props} (1,2),
      \ref{lemma:allT-props} (1) and hypotheses $(\f{BST}\ (\Node\ y\ l\ r))$}
    & 
    \begin{cases}
      (\insert\ x\ (\BST2list\ l))\app[y]\app(\BST2list\ r)
      & \text{if $x<y$,}\\
      (\BST2list\ l)\app[y]\app(\insert\ x\ (\BST2list\ r))
      & \text{if $x\geq y$,}\\
    \end{cases} \\
    \stepi{induction hypotheses}
    & 
    \begin{cases}
     (\BST2list\ (\istBST\ x\ l))\app[y]\app(\BST2list\ r)
      & \text{if $x<y$,}\\
      (\BST2list\ l)\app[y]\app(\BST2list\ (\istBST\ x\ r))
      & \text{if $x\geq y$,}\\
    \end{cases} \\
    \stepi{def. $\BST2list$}
    & \begin{cases}
      \BST2list\ (\Node\ y\ (\istBST\ x\ l)\ r)
      &\text{if $x<y$}, \\
      \BST2list\ (\Node\ y\ l\ (\istBST\ x\ r))
      &\text{if $x\geq y$}. \\
      \end{cases} \\
    \stepi{def. $\istBST$}
    & \BST2list\ (\istBST\ x\ (\Node\ y\ l\ r))
 \end{array}
\]
\end{proof}

Again, we note that the insertion function preserves the invariant.

\begin{proposition}
For every value $x$ and tree $t$,
\[ (\f{BST}\ t) \Rightarrow (\f{BST}\ (\istBST\ x\ t))). \]
\end{proposition}
\begin{proof}
  Induction on $t$. The base case follows immediately from the
  definitions. For the induction step we have:
\[
  \begin{array}{rl}
    & (\f{BST}\ (\istBST\ x\ (\Node\ y\ l\ r))) \\
    \stepe{def. $\istBST$}
    & 
    \begin{cases}
      (\f{BST}\ (\Node\ y\ (\istBST\ x\ l)\ r))
      & \text{if $x<y$,}\\
      (\f{BST}\ (\Node\ y\ l\ (\istBST\ x\ r)))
      & \text{if $x\geq y$,}\\
    \end{cases} \end{array}
\]
\noindent In fact, when $x<y$ we have:
\[
\begin{cases}
  (\AllT\ (<y)\ (\istBST\ x\ l))
  &\text{by lemma \ref{lemma:allT-props} (2)} \\
  (\AllT\ (\geq y)\ r)
  &\text{by $(\f{BST}\ (\Node\ y\ l\ r))$} \\
  (\f{BST}\ (\istBST\ x\ l))
  &\text{by induction hypotheses and $(\f{BST}\ (\Node\ y\ l\ r))$} \\
  (\f{BST}\ r)
  &\text{by $(\f{BST}\ (\Node\ y\ l\ r))$}
\end{cases}
\]
\noindent We reason similarly when $x\geq y$.
\end{proof}

And the required result follows directly by induction.

\begin{corollary}
\label{prop:buildbst}
For every list $l$,
\[ (\f{BST}\ (\buildBST\ l)) \]
\end{corollary}
\begin{proof}
  Simple induction on $l$.
\end{proof}

\newpage

\section{Alternative Derivation}
\label{app:sec:altern-deriv}

In this appendix we present a slight variation on the strategy for
deriving the sorting algorithms. This variation clarifies the role of
the invariants on intermediate structures in the correctness argument
of these algorithms.

When we compare the proof effort required to establish the correctness
of the ``sorting by insertion'' algorithms, we note that there is
significant difference between $\f{isortLT}$ and the other two
algorithms ($\isortH$ and $\isortBST$). As explained in the main text,
this is because the correctness for the last two algorithms depend on
properties of the intermediate structure. However, we can explain that
difference at a more abstract level --- one might argue that
$\f{isortLT}$ is closer to the specification of a \emph{generic
  insertion sort} presented at Section \ref{sec:sori}. To illustrate
this point, let us recall the definition of these algorithms (we omit
the definitions not relevant for this discussion):
\[
\begin{array}{l}\\
\begin{array}{lcl}
\f{isortLT}  & = & \f{LT2list} \after \f{buildLT}\\ 
\isortH  & = & \H2list \after \buildH\\ 
\isortBST  & = & \BST2list \after \buildBST\\
\f{LT2list} & = & \f{foldLTree}\ (\odot)\ t\\
&& \mbox{ where } 
\begin{array}[t]{l}
t\ \f{Nothing} = [\ ]\\
t\ (\f{Just}\ x) = [x]
\end{array}\\
\H2list & = & \foldr\ \f{aux}\ \Nil \\ 
&& \mbox{where } \f{aux}\ x\ l\ r = x:(l \odot r)\\
\BST2list & = & \f{foldBTree}\ \f{aux}\ \Nil\\
&& \mbox{ where } \f{aux}\ x\ l\ r = l \app (x:r)
\end{array}
\end{array}
\]
\noindent We observe that $\f{LT2list}$ uses only $\odot$ to construct
(non trivial) lists. On the other side, $\H2list$ and $\BST2list$ make
use of other functions (namely $(:)$ and $(\app)$). That distinction
makes the later two sensible to the ordering attributes of the
intermediate tree.

Let us make one step back and define the following variants of
$\isortH$ and $\isortBST$ algorithms: \def\BT2list{\f{BT2list}}
\[
\begin{array}{l}\\
\begin{array}{lcl}
\isortH'  & = & \BT2list \after \buildH\\ 
\isortBST'  & = & \BT2list \after \buildBST\\ 
\BT2list & = & \foldr\ \f{aux}\ \Nil \\ 
&& \mbox{where } \f{aux}\ x\ l\ r = [x]\odot (l \odot r)\\
\end{array}
\end{array}
\]
Now, the conversion of binary trees into lists ($\BT2list$) does not assume any
ordering constrains on these trees. In fact, $\BT2list$ and $\f{LT2list}$
should be read as two instances of the same polytypic function.

It is interesting to verify that, for these modified functions, the
correctness argument is essentially the same as for $\f{isortLT}$.

\begin{proposition}
  $\isortH'$ and $\isortBST'$ are sort algorithms.
\end{proposition}
\begin{proof}
  We instantiate eqs.~(\ref{eq:tl1}) and~(\ref{eq:tl2}) for both
  functions. We set $\epsilon = \Empty$, and thus eq.~(\ref{eq:tl1})
  results directly from the definition. For eq.~(\ref{eq:tl3}), we
  need to prove that for every binary tree $t$ and value $x$,
\begin{align*}
(\BT2list \after (\istH\ x))\ t & = ((\insert\ x)
  \after \BT2list)\ t \\
(\BT2list \after (\istBST\ x))\ t & = ((\insert\ x)
  \after \BT2list)\ t \\
\end{align*}
These are proved by induction on the structure of $t$. We show the
proof of the first one (the second is similar). The base case is
trivial. For the induction step we have:
\[
  \begin{array}{rl}
    & \BT2list(\istH\ x\ (\Node\ y\ l\ r)) \\
    \stepi{def. $\istH$}
    & 
    \begin{cases}
      (\BT2list(\Node\ x\ (\istH\ y\ r)\ l))
      & \text{if $x<y$,}\\
      (\BT2list(\Node\ y\ (\istH\ x\ r)\ l))
      & \text{if $x\geq y$,}\\
    \end{cases} \\
    \stepi{def. $\BT2list$}
    & 
    \begin{cases}
      [x]\odot ((\BT2list\ (\istH\ y\ r))\odot(\BT2list\ l))
      & \text{if $x<y$,}\\
      [y]\odot ((\BT2list\ (\istH\ x\ r))\odot(\BT2list\ l))
      & \text{if $x\geq y$,}\\
    \end{cases} \\
    \stepi{induction hypotheses}
    & 
    \begin{cases}
      [x]\odot (([y]\odot(\BT2list\ r))\odot(\BT2list\ l))
      & \text{if $x<y$,}\\
      [y]\odot (([x]\odot(\BT2list\ r))\odot(\BT2list\ l))
    \end{cases} \\
    \stepi{comutativity and associativity of $\odot$}
    & 
    [x]\odot ([y]\odot((\BT2list\ l)\odot(\BT2list\ r))) \\
    \stepi{def. $\BT2list$}
    & 
    [x]\odot (\BT2list\ (\Node\ y\ l\ r))
  \end{array}
\]
\end{proof}

\medskip

In order to refine $\isortH'$ and $\isortBST'$ to heap sort and
quicksort, we should now proceed in two independent paths:
\begin{itemize}
\item to show that the construction of the intermediate tree can
  be performed co-inductively (i.e. $\buildH$ and $\buildBST$ are
  equal to $\f{unfoldhsort}$ and $\f{unfoldqsort}$ respectively);
\item to show that the tree conversion into the resultant list can be
  simplified to their standard formulation (i.e. $\BT2list$ can be
  replaced by $\H2list$ for the heapsort and by $\f{BST2list}$ for the
  quicksort).
\end{itemize}

The first point was performed in the main text (c.f. Propositions
\ref{prop:buildH-unfold} and \ref{prop:buildBST-unfold}). The second
is the one that should consider the ordering properties induced by the
building process for each case --- more precisely, one proves:
\begin{align*}
\BT2list \after \buildH & = \H2list \after \buildH \\
\BT2list \after \buildBST & = \BST2list \after \buildBST
\end{align*}
As in appendix \ref{app:sec:tree-invs}, it is convenient to make
explicit the structural invariants possessed by the intermediate
structures in each case. That is,
\begin{align*}
(\f{HEAP}\ t) \qquad\Longrightarrow\qquad \BT2list\ t & = \H2list\ t \\
(\f{BST}\ t) \qquad\Longrightarrow\qquad \BT2list\ t & = \BST2list\ t
\end{align*}

The proof require a simple lemma relating $\odot$ with ordering predicates.

\begin{lemma}
\label{lemma:facts-odot}
  For every
  \begin{enumerate}
\item $(\AllL\ (x\leq)\ l_1) \quad\Longrightarrow\quad [x]\odot l_1 =
    x:l_1 $
\item $(\AllL\ (x>)\ l_1) \quad\Longrightarrow\quad l_1\odot (x:l_2)
    = l_1 \app (x:l_2)$
\item $(\AllL\ p\ l_1) \wedge (\AllL\ p\ l_2)
  \quad\Longrightarrow\quad (\AllL\ p\ (l_1\odot l_2))$
  \end{enumerate}
\end{lemma}
\begin{proof}
  The first two are proved by simple induction on the structure of
  $l_1$. The third by mutual induction on $l_1$ and $l_2$.
\end{proof}

Now, the required properties follow by simple induction. The base case is, in
both cases, trivial. For the induction step, we have for $\H2list$:
\[
  \begin{array}{rl}
    & \BT2list\ (\Node\ x\ l\ r)) \\
    \stepi{def. $\BT2list$}
    & [x]\odot((\BT2list\ l)\odot(\BT2list\ r)) \\
    \stepi{induction hypotheses}
    & [x]\odot((\H2list\ l)\odot(\H2list\ r)) \\
    \stepi{def. of $\f{HEAP}$ and lemma \ref{lemma:facts-odot} (1,3)}
    & x:((\H2list\ l)\odot(\H2list\ r)) \\
    \stepi{def. $\H2list$}
    & \H2list\ (Node\ x\ l\ r)
 \end{array}
\]
\noindent and for $\BST2list$:
\[
  \begin{array}{rl}
    & \BT2list\ (\Node\ x\ l\ r)) \\
    \stepi{def. $\BT2list$}
    & [x]\odot((\BT2list\ l)\odot(\BT2list\ r)) \\
    \stepi{induction hypotheses}
    & [x]\odot((\BST2list\ l)\odot(\BST2list\ r)) \\
    \stepi{commutativity and associativity of $\odot$}
    & (BST2list\ l)\odot([x]\odot(\BST2list\ r)) \\
    \stepi{def. of $\f{BST}$ and lemma \ref{lemma:facts-odot} (1,2)}
    & (\BST2list\ l)\odot(x:(\BST2list\ r)) \\
    \stepi{def. $\BST2list$}
    & \BST2list\ (Node\ x\ l\ r)
 \end{array}
\]

\end{document}